# Freezing of a Disorder Induced Quantum Spin Liquid

Xiao Hu[1], Daniel M. Pajerowski[2], Depei Zhang[1,+], Andrey A. Podlesnyak[2], Yiming Qiu[3], Qing Huang[4], Haidong Zhou[4], Israel Klich[1], Alexander I. Kolesnikov[2], Matthew B. Stone[2], and Seung-Hun Lee[1,*]

[1] Department of Physics, University of Virginia, Charlottesville, Virginia 22904, USA.

[2] Neutron Scattering Division, Oak Ridge National Laboratory, Oak Ridge, Tennessee 37831, USA.

[3] NIST Center for Neutron Research, National Institute of Standards and Technology, Gaithersburg, Maryland 20899, USA.

[4] Department of Physics and Astronomy, University of Tennessee, Knoxville, Tennessee 37996, USA.

+ Now at the Neutron Scattering Division, Oak Ridge National Laboratory, Oak Ridge, Tennessee 37831, USA.

* Corresponding Author: Seung-Hun Lee

Email: shlee@virginia.edu

## Abstract

Sr$_2$CuTe$_{0.5}$W$_{0.5}$O$_6$ is a square-lattice magnet with super-exchange between $S = \frac{1}{2}$ Cu$^{2+}$ spins mediated by randomly distributed Te and W ions. Here, using sub-K temperature and 20 µeV energy resolution neutron scattering experiments we show that this system transits from a gapless disorder-induced quantum spin liquid to a new quantum state below $T_f = 1.7(1)$ K, exhibiting a weak frozen moment of $\frac{\langle S \rangle}{S} \sim 0.1$ and low energy dynamic susceptibility, $\chi''(\hbar\omega)$, linear in energy which is surprising for such a weak freezing in this highly fluctuating quantum regime.

One of the most intriguing and elusive states of matter are quantum spin liquids (QSLs). QSLs are non-trivial, highly entangled states. Their unique features include spin and charge fractionalization and they have a possible application to topological quantum computing. Such states, envisioned and coined in Anderson's paper [1], exhibit strong zero-point fluctuations that prevent conventional magnetic long-range order. Since this first description, the subject grew in importance due to suggestions of a relation to superconductivity [2, 3], the fractional quantum hall effect [4] and topological order [5, 6]. From the late 1980s, experimental efforts to find QSLs in nature have been mainly focused on geometrically frustrated magnets since in such systems spins cannot order due to competing interactions [7]. Theoretical interest in QSLs has also grown substantially after the establishment of stable QSLs in exactly solvable lattice models such as Kitaev's honeycomb model [8]. The remarkable body of works on the subject are detailed in a number of reviews [9, 10]. While promising candidates for QSLs behavior have been geometrically frustrated quantum magnets [11-15], it has been also suggested that disorder in frustrated magnets can also stabilize "glassy" QSL phases [16-19], further expanding theoretical efforts [18, 20-23] and the experimental search for QSLs into disordered materials [17, 24-26].

Despite the huge efforts over more than 30 years, an experimental identification of a material with a true gapless quantum spin liquid ground state has not been conclusively established. A crucial stumbling block for experimentally establishing a quantum spin liquid is that many quantum spin liquid candidates freeze at very low temperatures. For example, α-RuCl$_3$ orders at 7 K [12] and NiGa$_2$S$_4$ freezes below 8.5 K [27]. Theoretically, it has been shown in some models that local coupling to the environment may result in local freezing leading to glassy behavior of quantum spin liquid models. Local freezing into metastable states can be favored entropically over remaining in the QSL state. It is unknown whether such freezing is a salient feature to be expected of typical spin liquid candidates. Such systems are considered to be in the spin liquid state for temperatures below the Curie-Weiss temperature, $\Theta_{CW}$, however it is an open question whether as the temperature approaches absolute zero they will exhibit some freezing or remain in the spin liquid state. A major experimental difficulty of approaching this regime is that most QSL candidates have small values of $|\Theta_{CW}|$, making observation of potential spin freezing experimentally inaccessible.

$Sr_2CuTe_{0.5}W_{0.5}O_6$ offers a good model system to search for a disorder-induced QSL. In this system the quantum spins of $Cu^{2+}$ ions form a two-dimensional square lattice. Despite the presence of strong antiferromagnetic interactions indicated by the Curie-Weiss temperature of -71 K [26], disorder in the magnetic couplings induced by random distribution of Te and W ions suppresses magnetic order. Previous experiments did not observe freezing at temperatures well below $|\Theta_{CW}|$. In particular, a muon spin-relaxation study reported that the spin correlations remain entirely dynamic down to 19 mK [26]. Interestingly, neutron scattering studies show that the dynamic spin correlations in the QSL phase exhibit collective spin wave-like excitations over the entire magnetic band [28]. Furthermore, the specific heat measurements exhibit an anomaly where a nearly linear in $T$ specific heat below ~ 7 K changes, upon further cooling, into quadratic behavior with $T$ below ~ 1.5 K [24]. This quadratic behavior may signal a transition from a QSL into a frozen state.

These seemingly contradicting experimental results call for a careful experimental study to understand the low temperature properties of this two-dimensional QSL candidate.

Here, we report experimental evidence for freezing and existence of Goldstone-like modes in Sr$_2$CuTe$_{0.5}$W$_{0.5}$O$_6$ at low temperatures. Experimental evidence comes from time-of-flight neutron scattering measurements with an extreme energy resolution of the half-width-at-half-maximum (HWHM) of 20 μeV down to sub-K temperatures. Our results show that this system exhibits some freezing below $T_f \sim 1.7(1)$ K even though the frozen moment is very small, $\langle S \rangle/S \sim 0.1$. Below $T_f$ the imaginary part of the dynamical susceptibility, $\chi''(\hbar\omega)$, behaves linearly with energy transfer, $\hbar\omega$, for $\hbar\omega < k_B T_f$, where $k_B$ is the Boltzmann constant, with the characteristic spin relaxation rate increasing with decreasing temperatures. The spatial spin correlations are two-dimensional and short-range with an in-plane correlation length of $\xi = 8.4(9)$ Å $\sim \sqrt{2}\, d_{NN}$, where $d_{NN}$ is the distance between nearest neighbor spins. In the quantum spin liquid state above $T_f$, the spatial spin correlations have the same nature at low energies as in the frozen state, i.e., short-range and two-dimensional. On the other hand, $\chi''(\hbar\omega)$ behaves as $\tan^{-1}\left(\frac{\hbar\omega}{\Gamma_{min}}\right)$ at low energies indicating the presence of a distribution of the spin relaxation rate with the lower limit energy $\Gamma_{min}$. $\Gamma_{min}$ behaves as a power law with temperature, $\Gamma_{min}/|J| = \left(\frac{k_B T}{|J|}\right)^\alpha$, with $\alpha = 1.3(1)$. These results tell us that Sr$_2$CuTe$_{0.5}$W$_{0.5}$O$_6$ transits from a gapless disorder-induced quantum spin liquid to a new quantum state below $\sim 1.7(1)$ K, exhibiting a weak frozen moment and low energy dynamic susceptibility that is linear in energy consistent with Halperin-Saslow-like excitations.

The time-of-flight (TOF) neutron scattering experiments were performed on an 8 g powder sample of $Sr_2CuTe_{0.5}W_{0.5}O_6$. To examine the magnetic excitations in the entire energy range, we performed the experiments with three different neutron incident energies, $E_i$, using two different spectrometers at the Spallation Neutron Source (SNS) located at the Oak Ridge National Laboratory; $E_i$ = 1.55 and 3.32 meV were used at the Cold Neutron Chopper Spectrometer (CNCS) [29] to focus on low energy excitations, and $E_i$ = 45 meV at the Fine-Resolution Fermi Chopper Spectrometer (SEQUOIA) [30] to probe high energy excitations up to the top of the magnetic energy band. Here we stress that the experiments with $E_i$ = 1.55 meV provided the instrumental energy resolution of 20 µeV HWHM which corresponds to 0.23 K. Such excellent energy resolution coupled with sub-K measurement temperatures are essential to investigate the true ground state of this magnetic system.

Fig. 1 (a) shows color contour maps of the inelastic neutron scattering cross section, $S(Q,\hbar\omega)$, as a function of momentum transfer, $Q$, and $\hbar\omega$, acquired at 0.25 K with $E_i$ = 1.55 and 3.32 meV, and at 5 K with $E_i$ = 45 meV. The $E_i$ = 45 meV data are shown only down to ~ 3 meV which corresponds to ~ 35 K and the magnetic excitations above 3 meV are expected to be similar at 5 K and 0.25 K. The figure shows that the magnetic excitations extend from at least 0.05 meV up to 20 meV. The intensity was normalized to obtain $S(Q,\hbar\omega)$ in an absolute unit of 1/meV/$Cu^{2+}$ by comparing the nuclear Bragg reflections to the calculated nuclear structure factors [31, 32]. The total signal from 0.05 to 20 meV, $\frac{\int_{BZ}\int_{0.05\,meV}^{20\,meV} S(Q,\hbar\omega)/[f(Q)]^2 d(\hbar\omega)dQ}{\int_{BZ} dQ}$, where $f(Q)$ is the $Cu^{2+}$ magnetic form factor, was estimated to be 0.5(1) which is consistent with the sum rule for the isotropic quantum spin of $\frac{2}{3}S(S+1) = 0.5$.

As shown in the Fig. 1 (a) and 1 (b), for $\hbar\omega \lesssim 7$ meV the magnetic excitations exhibit a prominent peak at $Q \sim 0.6$ Å$^{-1}$. On the other hand, for $\hbar\omega \gtrsim 7$ meV the magnetic excitations are almost featureless in $Q$, which is due to the Van Hove singularity of the top of the magnetic energy band from a powder sample. These overall features of $S(Q, \hbar\omega)$ can be understood as being due to a powder-averaged spin wave spectrum in a long range ordered magnetic state, similarly to the $S(Q, \hbar\omega)$ reported for the two mother compounds $Sr_2CuTeO_6$ [33, 34] and $Sr_2CuWO_6$ [35, 36], both of which exhibit long range order long range at low temperatures even though their ordered magnetic structures are different.

If we closely examine the data of $Sr_2CuTe_{0.5}W_{0.5}O_6$, however, we notice a peculiar feature. Fig. 1 (b) shows the constant $\hbar\omega$ cuts of $S(Q, \hbar\omega)$, $S(Q)$, for four different energy ranges. For $\hbar\omega \gtrsim 0.8$ meV, $S(Q)$ shows long ranged spin-wave-like features as discussed above, exhibiting a prominent peak at $Q \sim 0.6$ Å$^{-1}$ for $\hbar\omega \lesssim 7$ meV and being almost featureless for $\hbar\omega \gtrsim 7$ meV. Note that $S(Q)$ for both $3 \leq \hbar\omega \leq 5$ meV (blue triangles) and $0.8 \leq \hbar\omega \leq 2$ meV (orange circles) are more or less symmetric about $Q \sim 0.6$ Å$^{-1}$. On the other hand, the $S(Q)$ for $0.05 \leq \hbar\omega \leq 0.8$ meV (black squares) is strikingly asymmetric in $Q$. This indicates that the very low energy spin fluctuations of $Sr_2CuTe_{0.5}W_{0.5}O_6$ are due to low-dimensional dynamic spin fluctuations. The dimensional crossover of the spin fluctuations from being three-dimensional to lower-dimensional and the low-dimensional low-energy spin fluctuations may hold a key in understanding the anomalous low temperature magnetic properties that previous studies reported for this system [24-26]. The focus of this paper is how the low energy excitations below $\hbar\omega \sim 0.8$ meV evolve as a function of temperature down to sub-K.

The TOF measurements with $E_i = 1.55$ meV were performed at nine different temperatures from 0.25 K to 12 K. Among them, Fig. 2 shows the $S(Q, \hbar\omega)$ for four temperatures, 7, 4, 1.7, and 0.25 K. At all temperatures, the low energy excitations are dominated by the gapless streak that is centered at $Q \sim 0.6$ Å$^{-1}$. The measured scattering intensity indicates that the system is gapless down to the lowest energy $\hbar\omega \sim 0.05$ meV that can be accessed by the instrument energy resolution. Upon cooling from 7 K to 1.7 K, $S(Q, \hbar\omega)$ increases in strength for low energies of $\hbar\omega < 0.3$ meV, and the spectral weight of $S(Q, \hbar\omega)$ gradually shifts to lower energies. Surprisingly, however, upon further cooling from 1.7 K to 0.25 K, the strong low energy spin fluctuations below 0.2 meV become weak as shown as the red squire symbols in Fig. 3 (a). The depletion of $S(Q, \hbar\omega)$ at low energies upon cooling is typically a signature of spin freezing or development of static order of spins.

In order to investigate how the spectral weight of the low energy spin fluctuations evolve upon cooling, we integrated the scattering cross section $\tilde{S} \equiv \int_{0.4 \text{ Å}^{-1}}^{1.0 \text{ Å}^{-1}} \int_{0.05 \text{ meV}}^{0.2 \text{ meV}} S(Q, \hbar\omega) / [f(Q)]^2 d(\hbar\omega) dQ$ over a range of $Q$ and $\hbar\omega$ covering the prominent low energy peak centered at $0.6$ Å$^{-1}$ and plotted it as a function of $T$. As shown as the red square symbols in Fig. 3 (b), as $T$ decreases from 12 K to 2 K, the low energy spin fluctuations $\tilde{S}$ get gradually stronger. When $T$ decreases further from 2 K, however, the low energy spin fluctuations gradually weaken, transferring to the elastic channel (see Fig. 3 (c)). This strongly indicates that the spins indeed freeze below $T_f \sim 1.7(1)$ K. These findings starkly contradict the previous muon spin relaxation (µSR) study that reported a spin liquid state down to 19 mK. How could the µSR measurements not be able to detect the spin freezing? The clue comes from the fact that the spin freezing is very weak: as shown in Fig. 3 (b), the frozen spectral weight is $\Delta \tilde{S} = \tilde{S}(2\text{ K}) - \tilde{S}(0.25\text{ K}) \cong 0.003$ in the unit of 1/Cu$^{2+}$. Thus, below $T_f$ only 0.6 % out of the total spectral weight of the isotropic quantum spin that is $\frac{2}{3}S(S+1) = 0.5$ is frozen and the rest is fluctuating.

To study the nature of the weak spin freezing, we plotted the elastic magnetic scattering cross section, $S_{elas}^{mag}(Q, 0.25\text{ K}) = \int_{-0.02\text{ meV}}^{0.02\text{ meV}} S(Q, \hbar\omega) d(\hbar\omega)$ measured at 0.25 K, after background subtraction. Here background was determined by averaging similar elastic $S_{elas}(Q)$ measured at three different temperatures 4 K, 7 K, and 12 K above $T_f$, to increase the statistics. As shown in Fig. 3 (c), $S_{elas}^{mag}(Q, 0.25\text{ K})$ exhibits an asymmetric broad peak at $Q \sim 0.6$ Å$^{-1}$ similarly to the low energy gapless excitations shown in Fig. 2. This implies that the static correlations of the frozen spins are basically the same as the dynamic correlations of the fluctuating moments. The black line is the fit to a phenomenological Lorentzian function with a two-dimensional correlation length of $\xi = 12(6)$ Å that will be explained in details later. The large error for $\xi$ is due to the weak signal and resulting poor statistics.

To investigate the nature of the critical spin fluctuations at low energies, we have generated constant-$Q$ and constant-$\hbar\omega$ cuts from $S(Q, \hbar\omega)$ taken at nine different temperatures spanning the spin freezing transition, some of which are shown in Fig. 2. $S(\hbar\omega)$ was then converted to the imaginary part of the dynamic susceptibility $\chi''(\hbar\omega)$ using the fluctuation dissipation theorem. The resulting $S(Q)$ and $\chi''(\hbar\omega)$ for four different temperatures are shown in Fig. 4 (a) and (b), respectively.

Firstly, note that $S(Q)$ exhibits a prominent asymmetric peak with a maximum at $Q \approx 0.6$ Å$^{-1}$ that corresponds to $Q = \left(\frac{1}{2}, 0, 0\right)$, a sharp edge at lower $Q$s, and a long tail at higher $Q$s. There is another peak at $Q \approx 1.3$ Å$^{-1}$ that corresponds to $Q = \left(\frac{1}{2}, 1, 0\right)$. Thus, the low energy spin fluctuations have a characteristic antiferromagnetic wavevector of $q_m = \left(\frac{1}{2}, 0, 0\right)$. For a quantitative analysis of the spin dynamical correlation, we fit $S(Q)$ to the product of the independent lattice-Lorentzian functions [32, 37],

$$\frac{d\sigma(\mathbf{Q})}{d\Omega} \propto |F_m^\perp(\mathbf{Q})|^2 \prod_\alpha \frac{\sinh(\xi_\alpha^{-1})}{\cosh(\xi_\alpha^{-1}) - \cos[(\mathbf{q}_m - \mathbf{Q}) \cdot \hat{\mathbf{r}}_\alpha]}. \tag{1}$$

Here $F_m^\perp(\mathbf{Q}) = f(Q)\Sigma_\nu \mathbf{M}_\nu^\perp e^{-i\mathbf{Q}\cdot\mathbf{r}_\nu}$, where $\mathbf{M}_\nu$ and $\mathbf{r}_\nu$ are the staggered magnetic moment and the position of a Cu$^{2+}$ ion at the site $\nu$, respectively, and $f(Q)$ is the Cu$^{2+}$ magnetic form factor. $\xi_\alpha$ and $\hat{\mathbf{r}}_\alpha$ are the spin-correlation length and the unit cell lattice vector along the crystallographic axis ($\alpha = a, b, c$), respectively. The scattering cross section was convoluted with the instrumental resolution to fit the data. In the fitting, we used two different correlation lengths, an isotropic in-plane correlation length, $\xi = \xi_a = \xi_b$, and an out-of-plane correlation length, $\xi_c$. The fitting results are shown as the color-coded solid lines in Fig. 4 (a). For all the low temperatures considered, the out-of-plane $\xi_c$ was negligible, confirming the two-dimensionality of the critical spin fluctuations. Furthermore, as shown as black triangles in Fig. 3 (d), the in-plane correlation length, $\xi$, is very short, $\xi = 8.4(9)$ Å $\sim \sqrt{2}\, d_{NN}$ at 0.25 K, in which $d_{NN}$ is the distance between the nearest neighboring Cu$^{2+}$ ions. And $\xi$ gets slightly shorter above $T_f$: $\xi = 7.1(8)$ Å at 12 K. Thus, the critical spin flutuations at low temperatures have very short two-dimensional correlations that fall off quickly when the distance between the quantum spins goes beyond the distance between the second nearest neighbors.

Fig. 4 (b) shows $\chi''(\hbar\omega)$. At 7 K that is far below the Curie-Weiss temperature of Sr$_2$CuTe$_{0.5}$W$_{0.5}$O$_6$, $|\Theta_{CW}| = 71$ K [26] and well above $T_f$, the system is in a spin liquid state. In this state, as shown as the orange squres, $\chi''(\hbar\omega)$ gradually increases with increasing $\hbar\omega$. Upon cooling down to 1.7 K $\approx T_f$, $\chi''(\hbar\omega)$ softens, i.e., the spectral weight gradually shifts to lower energies. This low energy behavior is expected for a spin liquid since the energy scale of the low energy spin fluctuations in a spin liquid is temperature, $k_B T$, where $k_B \approx 0.086$ meV/K is the Boltzmann constant. For a quantitative analysis of the low energy fluctuations, we compare $\chi''(\hbar\omega)$ to a phenomenological function, $\chi''(\hbar\omega) \propto \tan^{-1}\left(\frac{\hbar\omega}{\Gamma_{min}}\right)$, that assumes a broad distribution of spin relaxation rates with the lower limit of $\Gamma_{min}$ [38]. Fig. 3 (d) shows the resulting $\Gamma_{min}$ in a log scale as a function of $T$ (see red circles). The red line is a fit to a function, $\frac{\Gamma_{min}}{|J|} = \left(\frac{k_B T}{|J|}\right)^\alpha$, with an energy scale of the magnetic interactions, $|J| = 9(2)$ meV, and a power, $\alpha = 1.3(1)$. The value of $|J|$ being close to the previously-reported value of the dominant magnetic interaction in this system [28], $J_2 \sim -9$ meV, and $\alpha$ being close to 1 support our interpretation that Sr$_2$CuTe$_{0.5}$W$_{0.5}$O$_6$ is in a quantum spin liquid state above $T_f$. Upon further cooling below $T_f$, however, low-energy spin degrees of freedom get depleted in the frozen state (see Fig. 3 (b)) where

$\chi''(\hbar\omega) \propto \hbar\omega$ up to $k_B T_f \approx 0.15$ meV, as shown as the black squares and black dashed line in Fig. 4 (d). The linear behavior of $\chi''(\hbar\omega) \propto \hbar\omega$ for $\hbar\omega < k_B T_f$ is consistent with the quadratic behavior of specific heat [24], $C(T) \propto T^2$ for $k_B T < k_B T_f$.

The exploration of disorder induced QSLs brings with it new possibilities and challenges. In particular, disorder can induce magnetic frustration without the presence of geometric frustration, and can lead to new RG (renormalization group) fixed points featuring local excitations as well as unusual dynamical exponents [17-19, 39]. Moreover, disorder can both facilitate a liquid state as well as facilitate freezing. Here we uncover a freezing phenomenon of a disorder-induced quantum spin liquid in which, remarkably, development of an extremely small frozen moment induces substantial changes in low temperature properties. Our neutron scattering results showed that Sr$_2$CuTe$_{0.5}$W$_{0.5}$O$_6$ freezes below $T_f \sim 1.7(1)$ K. Our findings tell us that the previous reported specific heat ($C_v$) anomaly at 1.5 K [24] is indeed intrinsic to the system and it is due to the freezing transition. The frozen moment however is very small, $\langle S \rangle / S \sim 0.1$, and 99.4 % out of the total spectral weight is fluctuating. This is probably the reason why μSR measurements could not detect any freezing [26]. Despite the small frozen moment, the low energy behavior of $\chi''(\hbar\omega)$ changes through the transition. In the QSL state above $T_f$, $\chi''(\hbar\omega) \propto \tan^{-1}\left(\frac{\hbar\omega}{\Gamma_{min}}\right)$ with the characteristic low-bound relaxation rate $\Gamma_{min} \propto (k_B T)^\alpha$ with $\alpha = 1.3(1)$. In the weakly frozen state below $T_f$, $\chi''(\hbar\omega) \propto \hbar\omega$ for $\hbar\omega < k_B T_f$. These results are consistent with $C_v$ being linear just above $T_f$ and being quadratic below $T_f$ [24].

In several previous theoretical and experimental studies the state above $T_f$ was regarded as a valence-bond glass (VBG). The magnetic excitations however do not exhibit any singlet-to-triplet-excitations characteristic of valence-bonds (See Fig. 1 (a) and Ref. [1-5, 28]). Rather, the magnetic excitations at high energies resemble spin-wave excitations of the ordered state of $Sr_2CuWO_6$, even though the excitations are smeared in energy [28, 36]. Thus, we believe it is more appropriate to call the state of $Sr_2CuTe_{0.5}W_{0.5}O_6$ above $T_f$ a disorder-induced glassy quantum spin liquid rather than VBG. Below $T_f$, $\chi''(\hbar\omega) \propto \hbar\omega$, a possible explanation for which is the presence of two-dimensional Halperin-Saslow (HS) modes [40-42]. Such modes are the analogue of Goldstone modes, the collective massless excitations that are present when rotational invariance is broken. These imply that the frozen state is a quantum analogue of a spin jam state, a glassy state typical for non-dilute frustrated magnets. It is quite remarkable that such a weak frozen moment can support HS modes that seem to dominate the low temperature and low energy properties. This also presents a theoretical challenge to fully understand the mechanism of the freezing of the disorder-induced QSL and the nature of HS modes in this new highly fluctuating quantum regime.

## Acknowledgments

The work at the University of Virginia was supported by the U.S. Department of Energy, Office of Science, Office of Basic Energy Sciences under Award Number DE-SC0016144. A portion of this research used resources at the Spallation Neutron Source, a DOE Office of Science User Facility operated by the Oak Ridge National Laboratory. The work at University of Tennessee was supported by DOE under Award No. DE-SC-0020254. The work of I.K. was supported in part by the NSF grant DMR-1508245.

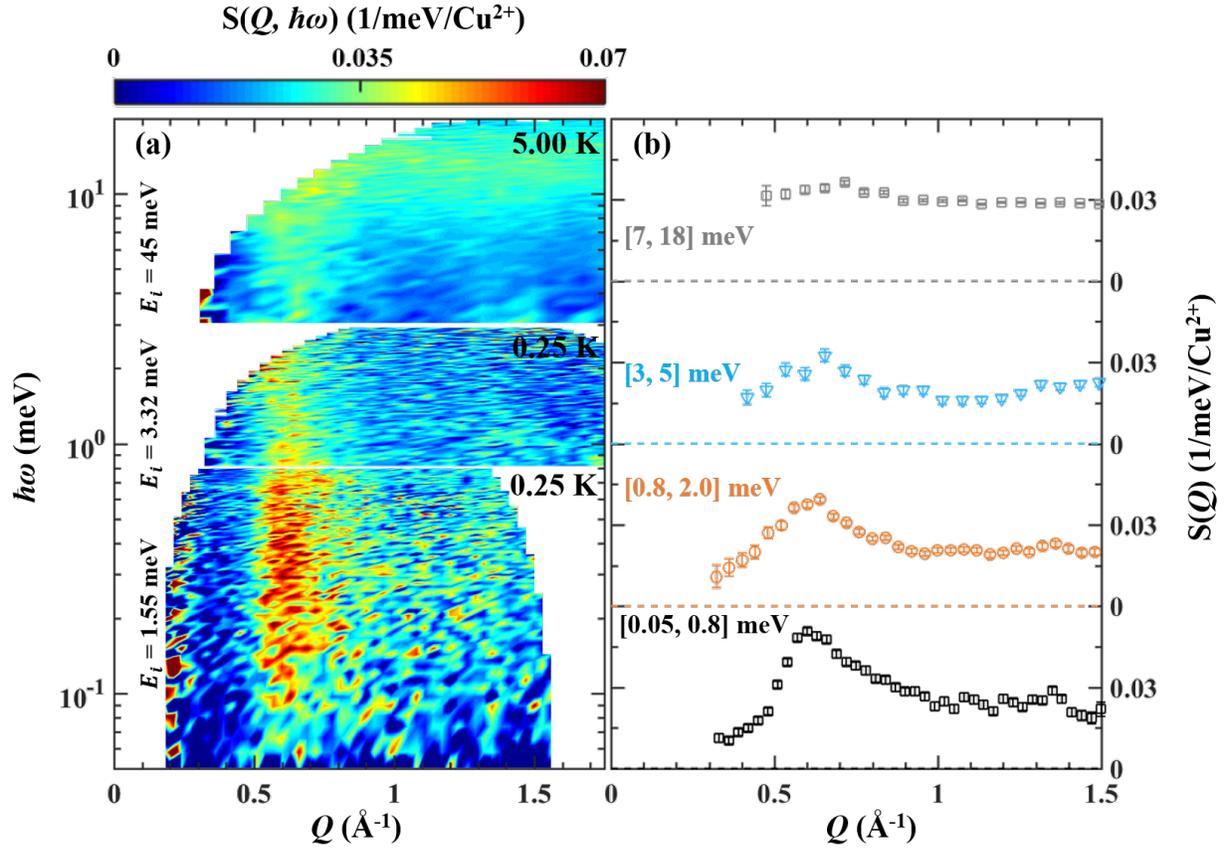

**Fig. 1. Inelastic neutron scattering data of $Sr_2CuTe_{0.5}W_{0.5}O_6$.** (a) Color contour maps of the neutron scattering intensity as a function of momentum transfer, $Q$, and energy transfer, $\hbar\omega$, measured with three different energies of incident neutrons, $E_i =$ 1.55 meV (bottom), 3.32 meV (middle), and 45 meV (top). The $E_i =$ 1.55 and 3.32 meV data were collected at the Cold Neutron Chopper Spectrometer (CNCS) and the $E_i =$ 45 meV data were collected at the Fine-Resolution Fermi Chopper Spectrometer (SEQUOIA) at SNS. (b) $Q$-dependence of the inelastic neutron scattering intensity, $S(Q) = \frac{\int S(Q,\hbar\omega)d(\hbar\omega)}{\int d(\hbar\omega)}$, for four different $\hbar\omega$-integration ranges. The dashed lines represent the zero value of the corresponding constant-$\hbar\omega$ cuts.

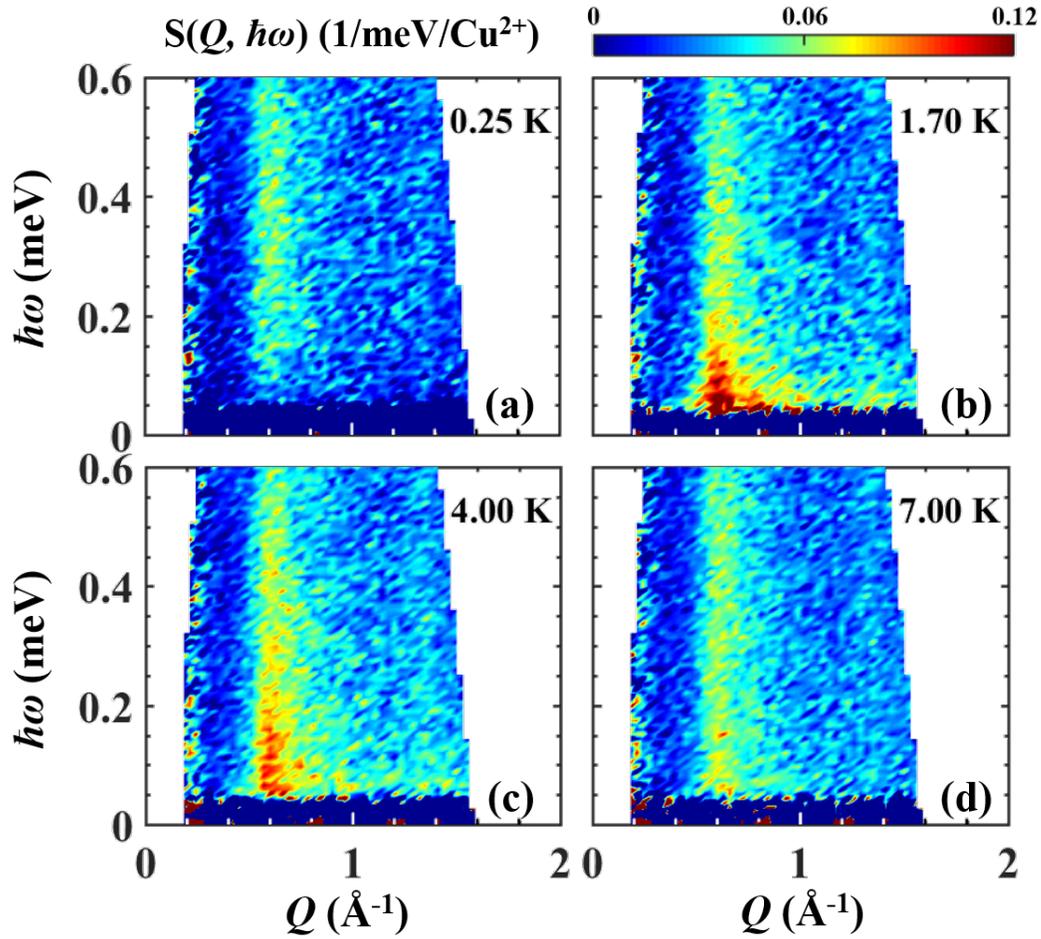

**Fig. 2. Low energy spin fluctuations of $Sr_2CuTe_{0.5}W_{0.5}O_6$.** Color contour maps of the low energy magnetic neutron scattering cross section, $S(Q, \hbar\omega)$, obtained with $E_i$ = 1.55 meV, measured at (a) 0.25 K, (b) 1.70 K, (c) 4.00 K, and (d) 7.00 K. The temperature independent background was determined by an algorithm using the detailed balance condition, $S(Q, -\hbar\omega) = e^{-\frac{\hbar\omega}{k_B T}} \cdot S(Q, \hbar\omega)$, and subtracted from the raw data to get $S(Q, \hbar\omega)$.

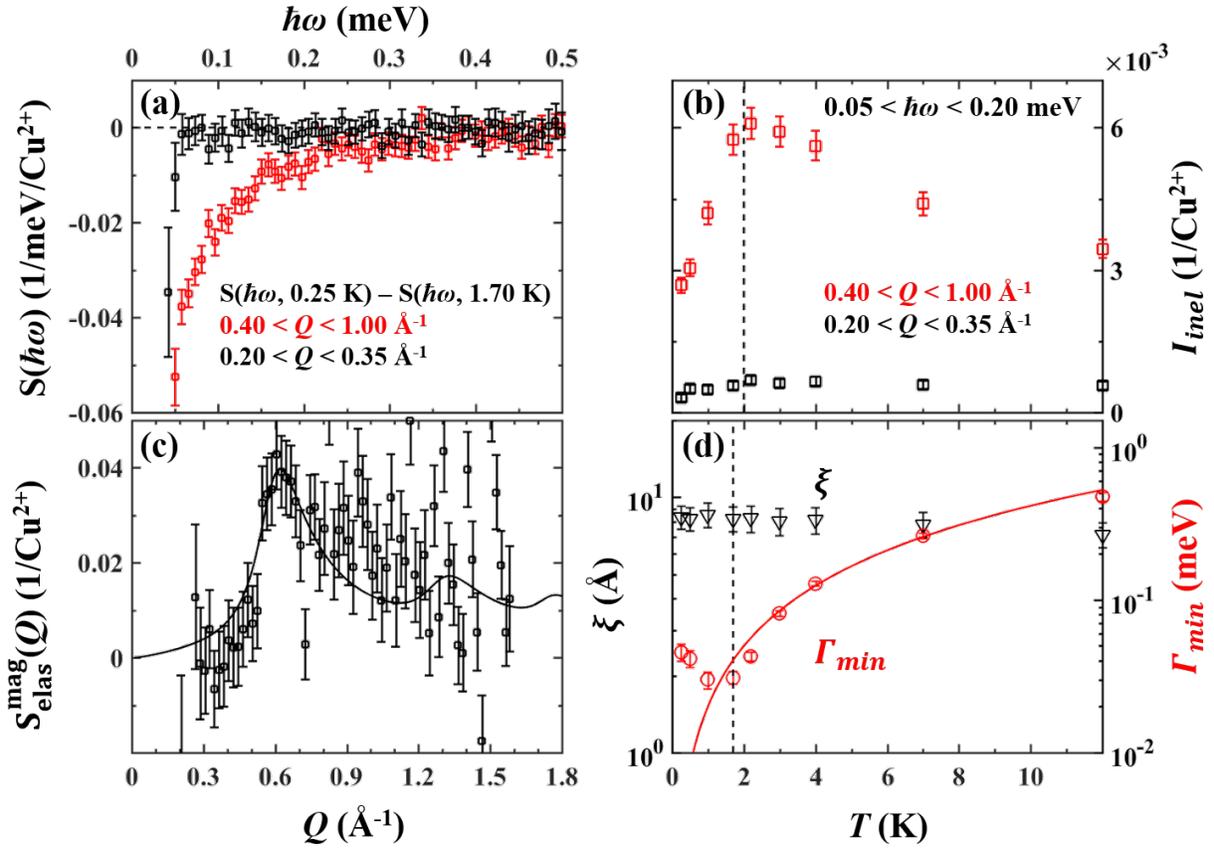

**Fig. 3.** (a) The $Q$-averaged neutron scattering intensity, $S(\hbar\omega) = \frac{\int S(Q,\hbar\omega)dQ}{\int dQ}$, obtained with $E_i$ = 1.55 meV at 0.25 K is shown after $S(\hbar\omega)$ at 1.7 K ~ $T_f$ was subtracted. The $Q$-integration range for the red squares was $0.4 \text{ Å}^{-1} < Q < 1.0 \text{ Å}^{-1}$ including $q_m = \left(\frac{1}{2}, 0, 0\right)$, while that for the black squares was $0.2 \text{ Å}^{-1} < Q < 0.35 \text{ Å}^{-1}$ below $q_m$. (b) The red squares represent the integrated spectral weight over an energy range of $0.05 < \hbar\omega < 0.20$ meV and a $Q$ range of $0.4 \text{ Å}^{-1} < Q < 1.0 \text{ Å}^{-1}$ as a function of temperature, $I_{inel} = \tilde{S} \equiv \int_{0.4 \text{ Å}^{-1}}^{1.0 \text{ Å}^{-1}} \int_{0.05 \text{ meV}}^{0.2 \text{ meV}} S(Q, \hbar\omega)/[f(Q)]^2 d(\hbar\omega)dQ$. The black squares represent $I_{inel}$ obtained with a $Q$ range of $0.2 \text{ Å}^{-1} < Q < 0.35 \text{ Å}^{-1}$ that show the $T$-independent backgrounds. (c) The elastic magnetic scattering cross section, $S_{elas}^{mag}(Q, 0.25 \text{ K}) = \int_{-0.02 \text{ meV}}^{0.02 \text{ meV}} S(Q, \hbar\omega)d(\hbar\omega)$ measured at 0.25 K, after background subtraction. Here background was determined by averaging similar elastic $S_{elas}(Q)$ measured at three

different temperatures 4 K, 7 K, and 12 K above $T_f$, to increase the statistics. (d) The red circles are the lower limit of the spin relaxation rates, $\Gamma_{min}$, extracted from the $y = y_0 \tan^{-1}\left(\frac{\hbar\omega}{\Gamma_{min}}\right)$ fitting of the imaginary part of the dynamics susceptibility, $\chi''(\hbar\omega)$, as a function of temperature. The red solid line represents the fitting results $\frac{\Gamma_{min}}{|J|} = \left(\frac{k_B T}{|J|}\right)^\alpha$ as discussed in the text. The black triangles are the correlation length, $\xi$, extracted from the $S(Q)$ fitting in Fig. 4(a).

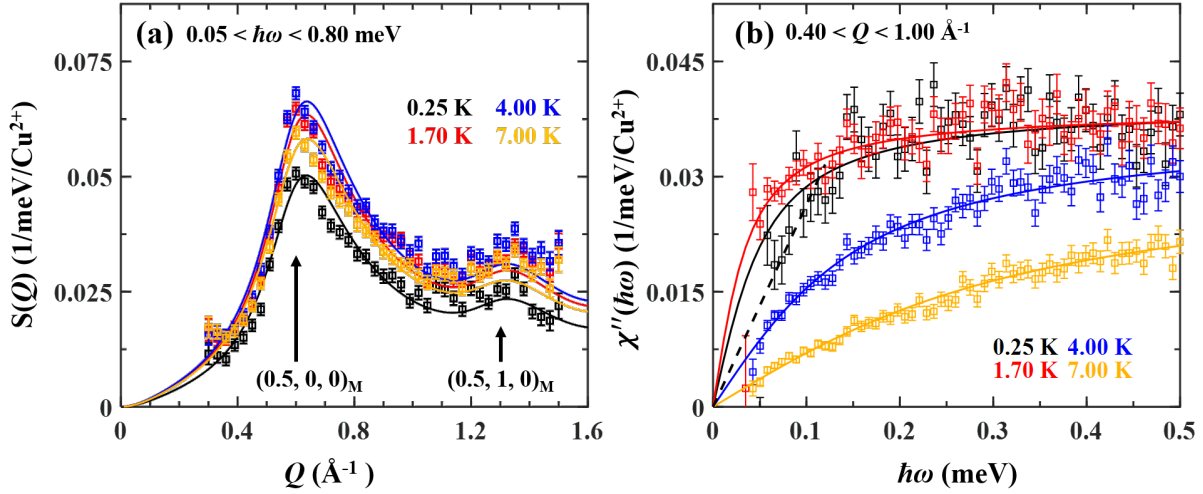

**Fig. 4.** $Q$ and $\hbar\omega$ dependences of low energy spin fluctuations in **$Sr_2CuTe_{0.5}W_{0.5}O_6$.** (a) $Q$-dependence of the low energy neutron scattering intensity obtained by averaging $S(Q,\hbar\omega)$ over $0.05 < \hbar\omega < 0.80$ meV, $S(Q) = \frac{\int_{0.05\,\text{meV}}^{0.80\,\text{meV}} S(Q,\hbar\omega)d(\hbar\omega)}{\int_{0.05\,\text{meV}}^{0.80\,\text{meV}} d(\hbar\omega)}$, at four different temperatures, 0.25 K (black squares), 1.70 K (red), 4.00 K (blue) and 7.00 K (orange). The color solid lines are the $S(Q)$ fitting results as discussed in the text. Two characteristic wave vectors, $q_m$ = (1/2, 0, 0) and (1/2, 1, 0), are indexed with black arrows. (b) The color squares are the energy dependence of the imaginary part of the dynamics susceptibility, $\chi''(\hbar\omega) = \left(1 - e^{-\frac{\hbar\omega}{k_B T}}\right) S(\hbar\omega)$, obtained by averaging and converting the inelastic neutron scattering intensity $S(Q,\hbar\omega)$ over $0.40 < Q < 1.00\,\text{Å}^{-1}$, $S(\hbar\omega) = \frac{\int_{0.40\,\text{Å}^{-1}}^{1.00\,\text{Å}^{-1}} S(Q,\hbar\omega)d(Q)}{\int_{0.40\,\text{Å}^{-1}}^{1.00\,\text{Å}^{-1}} d(Q)}$, at 0.25 K (black), 1.70 K (red), 4.00 K (blue), and 7.00 K (orange). The color solid lines represent fits by the spectral weight function of an arctangent type, $y = y_0 \tan^{-1}\left(\frac{\hbar\omega}{\Gamma_{min}}\right)$, which gives the lower bound ($\Gamma_{min}$) of the spin-relaxation rates. The black dashed line represents the linear $\hbar\omega$ dependence of low-energy fluctuations up to ~ 0.15 meV.